\documentclass[aps,prd,amssymb,eqsecnum,showpacs,nofootinbib]{revtex4}

\usepackage{amsmath}
\usepackage{graphics}

\allowdisplaybreaks

\newcommand{\be}{\begin{equation}}
\newcommand{\btheta}{\mbox{\boldmath$\theta$}}

\newcommand{\cpipii}{{\mathbf{p}}_1\times{\mathbf{p}}_2}
\newcommand{\csipi}{{\mathbf{S}}_1\times{\mathbf{p}}_1}
\newcommand{\csipii}{{\mathbf{S}}_1\times{\mathbf{p}}_2}
\newcommand{\cnsi}{{\mathbf{n}}_{12}\times{\mathbf{S}}_1}

\newcommand{\ee}{\end{equation}}

\newcommand{\md}{\mathrm{d}}
\newcommand{\mD}{\mathrm{D}}

\newcommand{\ncpi}{{\mathbf{n}}_{12}\times{\mathbf{p}}_1}
\newcommand{\ncpii}{{\mathbf{n}}_{12}\times{\mathbf{p}}_2}
\newcommand{\npi}{(\mathbf{n}_{12}\cdot\mathbf{p}_1)}
\newcommand{\npii}{({\bf n}_{12}\cdot{\bf p}_2)}
\newcommand{\nsi}{{\bf n}_{12}\times{\bf S}_1}
\newcommand{\nsii}{{\bf n}_{12}\times{\bf S}_2}

\newcommand{\piipii}{{\bf p}_2^2}

\newcommand{\pipi}{{\bf p}_1^2}
\newcommand{\pipii}{({\mathbf{p}}_1\cdot{\mathbf{p}}_2)}

\newcommand{\sapav}{\mathbf{S}_a\times\mathbf{p}_a}
\newcommand{\sbbfa}{{\bf S}_a^{\mathrm{c\,BBF}}}
\newcommand{\sbbfb}{{\bf S}_b^{\mathrm{c\,BBF}}}
\newcommand{\sbbfnca}{{\bf S}_a^{\mathrm{BBF}}}

\newcommand{\sipiv}{{\bf S}_1\times{\bf p}_1}
\newcommand{\siipiv}{{\bf S}_2\times{\bf p}_1}
\newcommand{\sipiiv}{{\bf S}_1\times{\bf p}_2}
\newcommand{\siipiiv}{{\bf S}_2\times{\bf p}_2}
\newcommand{\sinpi}{\big(S_1,n_{12},p_1\big)}
\newcommand{\sinpii}{\big(S_1,n_{12},p_2\big)}
\newcommand{\siinpi}{\big(S_2,n_{12},p_1\big)}
\newcommand{\siinpii}{\big(S_2,n_{12},p_2\big)}

\newcommand{\xpsa}{\mathbf{x}_a,\mathbf{p}_a,\mathbf{S}_a}
\newcommand{\xpsb}{\mathbf{x}_b,\mathbf{p}_b,\mathbf{S}_b}

\begin{document}

\title{Hamiltonian of two spinning compact bodies
with next-to-leading order gravitational spin-orbit coupling}

\author{Thibault Damour}
\email{damour@ihes.fr}
\affiliation{Institut des Hautes \'Etudes Scientifiques,
91440 Bures-sur-Yvette, France}

\author{Piotr Jaranowski}
\email{pio@alpha.uwb.edu.pl}
\affiliation{Faculty of Physics,
University of Bia{\l}ystok,
Lipowa 41, 15--424 Bia{\l}ystok, Poland}

\author{Gerhard Sch\"afer}
\email{gos@tpi.uni-jena.de}
\affiliation{Theoretisch-Physikalisches Institut,
Friedrich-Schiller-Universit\"at,
Max-Wien-Pl.\ 1, 07743 Jena, Germany}

\date{\today}

\begin{abstract}

A Hamiltonian formulation is given for the gravitational dynamics
of two spinning compact bodies to next-to-leading order
($G/c^4$ and $G^2/c^4$) in the spin-orbit interaction.
We use a novel approach (valid to linear order in the spins),
which starts from the second-post-Newtonian
metric (in ADM coordinates)  generated by two spinless bodies,
and computes the next-to-leading order precession, in this metric,
of suitably redefined ``constant-magnitude'' 3-dimensional
spin vectors ${\bf S}_1$, ${\bf S}_2$. We prove the Poincar\'e invariance of
our Hamiltonian by explicitly constructing  ten phase-space generators realizing
the Poincar\'e algebra. A remarkable feature of our approach is that it allows
one to derive the {\it orbital} equations of motion of spinning binaries
to next-to-leading order in spin-orbit coupling without having to
solve Einstein's field equations with a spin-dependent stress tensor.
We show that our Hamiltonian (orbital and spin) dynamics is
equivalent to the dynamics recently obtained by Faye, Blanchet, and
Buonanno, by solving Einstein's equations in harmonic coordinates.

\end{abstract}


\maketitle

\section{Introduction}

In view of the needs of upcoming gravitational-wave observations, it
is crucial to be able to describe in detail the dynamics of {\it spinning} 
compact binaries. We think that this aim will be fullfilled by combining the
knowledge acquired by analytical techniques with that obtained by numerical ones.
The present paper is devoted to a new, Hamiltonian analytical treatment
of the general relativistic dynamics of spinning binaries.

The dynamics of spinning bodies in general relativity is a rather complicated
problem which has been the subject of many works over many years (starting from
the pioneering contributions of Mathisson \cite{M37}, Papapetrou \cite{P51},
Pirani \cite{P56}, Tulczyjew \cite{T59}, and others).
This paper focusses on (gravitational) spin-orbit effects, i.e.\ dynamical effects
which are {\it linear} in the spins of a binary system. The spin-orbit
interaction can be analytically obtained as a post-Newtonian (PN) expansion.
The leading-order contribution of this expansion is proportional to $G/c^2$,
while the next-to-leading order  one contains two sorts of terms:
$G/c^4$ and $G^2/c^4$ (here $G$ denotes Newton's gravitational constant and
$c$ the speed of light). The first complete derivation of leading-order (LO)
spin-orbit effects in comparable-mass binary systems is due to
Barker and O'Connell \cite{Barker:1975ae,BOC79}. These authors
derived the spin-orbit interaction by considering the {\it quantum} scattering
amplitude of two spin-$\frac{1}{2}$ particles. This curious fact prompted several
authors to give purely {\it classical} derivations of LO spin-orbit effects
(see, e.g., Refs.\ \cite{DamourVarenna75,BornerEhlersRudolph,D'Eath:1975vw}).
For a discussion of LO spin-orbit effects in coalescing binary systems
see Refs.\ \cite{KWW93,Kidder95}.

The next-to-leading order (NLO) spin-orbit interaction was analytically
tackled only over the last few years. After a first incomplete attack due
to Tagoshi, Ohashi, and Owen \cite{TOO01},  complete results were obtained
very recently by Faye, Blanchet, and Buonanno \cite{FBB06} and
Blanchet, Buonanno, and Faye \cite{BBF06}. Reference \cite{FBB06}
calculated the translational equations of motion, as well as
the rotational equations of motion for compact spinning binaries to NLO
(as here, only terms linear in spin were considered). For their derivation,
Blanchet et al., working in harmonic coordinates, introduced the
pole-dipole energy-momentum tensor due to Tulczyjew \cite{T59} in the
Einstein field equations. They also  used the general-relativistic-covariant
{\em spin supplementary condition} (SSC) of Tulczyjew \cite{T59} or,
equivalently in the linear-in-spin approximation, of Pirani \cite{P56}.

The new derivation of NLO spin-orbit interactions in the present paper
is based on a novel approach, and is totally independent from the results
of Refs.~\cite{FBB06,BBF06}. At the end, we shall be able to connect
our results to those of \cite{FBB06,BBF06}, thereby giving us confidence
in the correctness of both investigations. We do not use Tulczyjew's
pole-dipole energy-momentum tensor. We do not either make use of the
Papapetrou (or, more completely, Mathisson-Papapetrou-Pirani) translational
equations of motion. Our starting point consists of the second post-Newtonian
(2PN) metric generated by spinless point masses in ADM coordinates, say
$g_\mathrm{(2PN)o}$. The crux of our approach then consists in noting that
(to linear order in the spins) it is enough to compute the NLO spin
precession equations in $g_\mathrm{(2PN)o}$ to derive the spin-orbit NLO
contribution in the Hamiltonian, say
$H^\mathrm{NLO}_\mathrm{so}({\bf x}_1,{\bf x}_2,{\bf p}_1,{\bf p}_2,{\bf S}_1,{\bf S}_2)$.
Then, from
$H^\mathrm{NLO}_\mathrm{so}({\bf x}_1,{\bf x}_2,{\bf p}_1,{\bf p}_2,{\bf S}_1,{\bf S}_2)$
we can derive the NLO spin-dependent terms in the translational equations
of motion (simply by using Hamilton's canonical evolution equations).
Technically, we shall derive the spin precession equations by starting from
the 4-dimensional parallel transport equation for the  spin 4-vector 
(with covariant spin supplementary condition), and then by rewriting them
in terms of a suitably defined 3-dimensional spin vector, having a constant
Euclidean magnitude. (This method is essentially that used in 
Ref.~\cite{DamourVarenna75} at the LO.) We shall then check the
Poincar\'e invariance of our Hamiltonian by explicitly constructing
ten phase-space generators realizing the Poincar\'e algebra (similarly
to the proof of the Poincar\'e invariance of the 3PN orbital Hamiltonian
given in \cite{DJS00}). After our construction, we shall give the
relation with the results obtained in Refs.~\cite{FBB06,BBF06}
in the form of explicit transformation formulae.

We leave to a sequent paper a discussion of the physical consequences of
our Hamiltonian formulation, and notably its use for improving the
description of spin effects within the {\em effective one-body approach} \cite{EOB}.

\section{3-dimensional Euclidean spin vector in curved spacetime,
and its angular velocity}

When working to linear order in the spin, the translational and
rotational equations of motion of a spinning particle in curved space
\cite{M37,P51,P56,T59} (see also \cite{MTW73} and \cite{FBB06}) 
read\footnote{In this paper, Greek indices run over the numbers
$0,1,2,3$, Latin indices over $1,2,3$, $\epsilon^{\lambda\rho\alpha\beta}$ is
the completely antisymmetric (flat spacetime) Levi-Civita symbol with
$\epsilon^{0123}=1$. Note that Ref.\ \cite{MTW73} uses an opposite sign convention
for $\epsilon^{\lambda\rho\alpha\beta}$, which leads to an opposite sign
on the right-hand-side of \eqref{eom1}.}
\begin{align}
\label{eom1}
m \frac{\mD u_\mu}{\md\tau}
&= \frac{1}{2} \frac{\epsilon^{\alpha\beta\lambda\rho}}{\sqrt{-g}}
\tilde{S}_\alpha u_\beta u_\nu R^\nu_{~\mu\lambda\rho},
\\[1ex]
\label{partra1}
\frac{\mD\tilde{S}_\mu}{\md\tau} &= 0,
\end{align}
where $u^\mu$ is the normalized 4-velocity of the spinning particle,
$u^\mu u_\mu = -1$, $m$ its conserved mass, and $\tilde{S}^\mu$ its
4-dimensional spin vector; in addition, $\tau$ denotes the proper time parameter,
${\md x^\mu}/{\md\tau}=cu^\mu$, $\mD$ the 4-dimensional
covariant derivative, $R^\mu_{~\nu\lambda\rho}$ the Riemann curvature
tensor, and $g$ the determinant of the 4-dimensional metric $g_{\mu\nu}$.

An important feature of our approach is that we shall not need to consider the
{\it translational}  equations of motion \eqref{eom1}. It will be enough to consider
the {\it rotational} ones \eqref{partra1}. One immediate consequence of \eqref{partra1}
is that the 4-dimensional length of $\tilde{S}^\mu$ is preserved
along the world line
\be
\label{4spin-constancy}
g^{\mu\nu}\tilde{S}_\mu\tilde{S}_\nu = s^2, \quad s^2 = \mbox{const},
\ee
where $g^{\mu\nu}g_{\nu\lambda}=\delta^{\mu}_{\lambda}$. The constant
scalar $s$ measures the proper magnitude of the spin.
The Eqs.\ \eqref{eom1} and \eqref{partra1}, to linear order in spin,
are compatible with the covariant SSC
\be
\label{ssc}
\tilde{S}_\mu u^\mu = 0.
\ee
At the same approximation, this (Pirani \cite{P56}) SSC is equivalent to
the Tulczyjew \cite{T59} one $S^{\mu \nu} p^{\rm kin}_\nu = 0$, where
$p^{\rm kin}_\mu = m c u_\mu + O(s^2)$ is the {\it kinematical} momentum (which differs from
the canonical momentum we shall use below), and where  $S^{\mu \nu}$ is the
antisymmetric spin tensor (see, e.g., \cite{FBB06}).

More explicitly,  Eq.\ \eqref{partra1} reads, when expressed in terms of
the coordinate time $t\equiv x^0/c$,
\be
\label{partra2}
\frac{\md\tilde{S}_\mu}{\md t}
= c\,{\Gamma^\nu}_{\mu\rho} \tilde{S}_\nu v^\rho,
\ee
where ${\Gamma^\nu}_{\mu\rho}$ are the Christoffel symbols
and $v^\mu\equiv c^{-1}\md x^\mu/\md t = u^\mu/u^0 = (1,v^i)$.
Note that, in this paper, we normalize the ``velocity''
$v^i\equiv c^{-1}\md x^i/\md t$
so that it is dimensionless.

In addition, we can use Eq.\ \eqref{ssc} to compute the covariant time
component of the spin vector in terms of its (covariant) spatial components:
\be
\label{ncssc}
\tilde{S}_0=-\tilde{S}_iv^i.
\ee
Substituting this result into Eq.\ \eqref{4spin-constancy} one finds that the
constancy of the 4-dimensional spin magnitude takes the 3-dimensional form
\be
G^{ij} \tilde{S}_i \tilde{S}_j = s^2,
\ee
where $G^{ij}$ is the symmetric matrix:
\be
\label{Gdef}
G^{ij} \equiv g^{ij} - g^{0i} v^j - g^{0j} v^i + g^{00} v^i v^j.
\ee
Now a technically very useful fact is that a {\em  positive-definite symmetric} 
matrix such as the one just defined, $G^{ij}$,
admits a {\em unique positive-definite symmetric square root}, say
$H^{ij} = H^{ji}$, such that
\be
\label{GHH}
G^{ij} = H^{ik}H^{kj}.
\ee
This uniqueness result (in some given coordinate system)  then naturally
leads us to {\em defining} a constant-in-magnitude {\it 3-dimensional Euclidean
spin vector} $ S_i \equiv S^i$ as\footnote{A slightly more geometrical way of
phrasing this definition would consist in saying that,
starting from a given coordinate system, we are constructing
a well-defined orthonormal ``rep\`ere mobile'' (or ``vierbein'')
along the worldline of a spinning particle,
with respect to which the covariant spin 4-vector has components $(0,S^i)$.
By definition, the spatial components of the metric
in this local orthonormal frame take the standard Euclidean values $\delta_{ij}$,
so that we can trivially raise or lower indices on our spin 3-vector.}
\be
\label{sic}
S_i \equiv H^{ij} \tilde{S}_j, \quad S_iS_i=s^2.
\ee

Upon further use of the spin supplementary condition \eqref{ncssc},
the spatial covariant component of the rotational equation of motion
\eqref{partra2} yields
\be
\label{dotsi}
\frac{\md\tilde{S}_i}{\md t} = \tilde{V}^{ij} \tilde{S}_j,
\ee
where
\be
\label{Vij}
\tilde{V}^{ij} \equiv c\Big( {\Gamma^j}_{i0} +
{\Gamma^j}_{ik}v^k - {\Gamma^0}_{i0}v^j -  {\Gamma^0}_{ik}v^jv^k
\Big).
\ee
Making use of Eqs.\ \eqref{sic} and \eqref{dotsi}
one can now easily derive an evolution equation for the constant-magnitude
3-dimensional spin vector $S_i$
(dot means differentiation with respect to the coordinate time $t$):
\be
\label{dotsic}
\dot{S}_i = V^{ij}S_j, \quad
V^{ij} \equiv \dot{H}^{ik} (H^{-1})^{kj}
+ H^{ik}\tilde{V}^{kl}(H^{-1})^{lj}.
\ee
The constancy of the Euclidean magnitude of $S_i$ implies that the
matrix $V^{ij}$ determining the ``rotational velocity'' of $S_i = S^i$
is {\it antisymmetric}: $V^{ij} = - V^{ji}$ (a result which is easily
checked to hold for the explicit expression of $V^{ij}$ given above).
It is then convenient to ``dualize'' $V^{ij}$ and to replace it by the
3-dimensional Euclidean (pseudo-) vector
\be
\label{defomega}
\Omega_i \equiv -\frac{1}{2} \varepsilon_{ijk} V^{jk}.
\ee
With this notation the rotational equation of motion  \eqref{dotsic} reads
\be
\label{Omega}
\dot{S}_i =  + \; \varepsilon_{ijk} \Omega_j S_k.
\ee
In other words, we get a Newtonian looking  {\it spin precession equation} 
$ \dot{\mathbf{S}} = \mathbf{\Omega} \times \mathbf{S}$. 

In summary, the angular velocity of rotation $\mathbf{\Omega}$
of the constant-magnitude spin 3-vector \eqref{sic}
is directly computable from the spacetime metric (and its Christoffel symbols)
by using the explicit formulas \eqref{Vij}, \eqref{dotsic}, and \eqref{defomega}.
(For the self-gravitating spinning particles we are considering, one will need,
as usual, to regularize the self-interaction terms hidden in the formal results
written above. See below.) Note that $\mathbf{\Omega}$ depends,
in general, both on the positions {\it and the velocities}
of all the particles in the system. Indeed, from the explicit formulas above,
one sees that $\mathbf{\Omega}$ depends on the velocity of the considered spinning
particle. Moreover, the metric and Christoffel symbols at the location of
some particle will depend on the positions and velocities of the {\it other} particles.

\section{Deriving the spin-orbit interaction Hamiltonian from the angular velocity
of the Euclidean spin 3-vector}

Let us now show how the knowledge of the just discussed spin angular
velocity vector $\mathbf{\Omega}$ allows one to derive the spin-orbit
interaction Hamiltonian $H_\mathrm{so}$, i.e.\ the part of the Hamiltonian
which is linear in the spin variables.

Let us first recall that a basic result in Hamiltonian dynamics
is Darboux's theorem which says that any (non singular)
symplectic form $\omega$ on an even-dimensional manifold
can always be (locally) rewritten (after a suitable
change of phase-space coordinates) in the canonical form
$\omega=\sum_A\md q^A \wedge \md p_A$. When considering $N$ (interacting)
spinning particles, the dimension of phase space is $N(3+3+2)=8N$,
because the description of each particle requires: 3 spatial coordinates,
3 momenta and {\it 2 spin degrees of freedom}, such as two angles
$ \theta, \phi$ needed to parametrize the direction of the (constant-magnitude)
spin 3-vector $S_i$. Darboux's theorem then means, in this case, that it is
always possible to redefine phase-space coordinates such that the symplectic form
takes the form
$$
\omega = \sum_a \Big( \sum_i \md q^i_a \wedge \md p_{ai}
+ s_a \md(-\cos\theta_a) \wedge \md\phi_a \Big).
$$
Here $a=1, \ldots ,N$ labels the various 
particles (with $N=2$ in our case), while $i=1,2,3$ labels the spatial dimensions.
We have written $\omega$ in the  form it is known to take in special relativity
\cite{Souriau, BelMartin}. In the latter case (and, say for simplicity, in the
case of free particles), the spin-dependent term in $\omega$
was shown to take (globally) the form indicated, with $s_a$ denoting the
magnitude of the conserved spin of the $a$th particle, in the sense of
\eqref{4spin-constancy}, and with $\theta_a$ and $ \phi_a$ denoting the polar
angles of the flat-space limit of the above-introduced constant-magnitude
Euclidean spin vector $S^i_a$, \eqref{sic}.
When considering the {\it interacting} case
(i.e.\ turning on a non-zero value of $G/c^2$), and when keeping, for simplicity,
only the terms linear in spin (so that one can expand the dynamics in powers of
both $G$ and $s_a$), it is easily checked
(by a perturbation analysis\footnote{I.e.\ by considering
general coordinate changes of the form $q' = q + O(s)$, $p' = p + O(s)$
and working linearly in the spins $s$.}) that it is always possible to construct
Darboux-type canonical coordinates where the spin degrees of freedom are
simply the polar angles (in a local orthonormal frame) of the above-introduced
constant-magnitude Euclidean spin vector $S^i_a$.\footnote{As we shall discuss below,
we can still modify $S^i_a$  by a rather general local rotation, but the
important point is that our definition of $S^i_a$, \eqref{sic}, is a smooth deformation
of the correct flat-spacetime limit.}

Finally, we can transcribe this result in the language of Poisson brackets
(instead of that of a symplectic form), by stating that there exist
phase-space variables $\mathbf{x}=(x^i_a)$, $\mathbf{p}=(p_i^a)$,
and $\mathbf{S}=(S_i^a)$ (with $a=1, \ldots ,N$, and $i=1,2,3$),
where $S_i^a$ are, say, the constant-magnitude vectors
\eqref{sic} such that the usual (Newtonian-like) Poisson brackets
\be
\label{pb}
\{x^i_a,p_j^b\} =\delta^b_a \delta^i_j,
\quad \{S_i^a,S_j^b\} = \delta^{a b}\varepsilon_{ijk}S_k^a,
\quad \mbox{zero otherwise},
\ee
apply to the case of a general-relativistically interacting sytem of 
$N$ spinning particles.

Note, however, that this result is essentially {\it kinematical}, and
has nearly  no {\it dynamical} content. To describe the dynamics
of interacting spinning particles, we need to  know the
expression of the Hamiltonian in terms of the canonical variables:
$H=H(\mathbf{x}_a,\mathbf{p}_a,\mathbf{S}_a)$. As we work linearly in the
spins, we look for an Hamiltonian of the general form:
\be
\label{hcomplete}
H({\bf x}_a,{\bf p}_a,{\bf S}_a)
= H_{\mathrm{o}}({\bf x}_a,{\bf p}_a)
+ H_{\mathrm{so}}({\bf x}_a,{\bf p}_a,{\bf S}_a).
\ee
Here, $H_\mathrm{o}$ denotes the orbital part of $H$,
while $H_\mathrm{so}$ contains all the linear-in-spin terms,
and can be called the ``spin-orbit part''.
The orbital Hamiltonian $H_{\mathrm{o}}$  is explicitly known up 
to the 3PN order \cite{DJS00,DJS01a}. Our aim here is to compute
the spin-orbit Hamiltonian $H_\mathrm{so}$ to NLO.
Because $H_\mathrm{so}$ is, by definition, linear in the spins we can
always write it in the general form
\be
\label{hsodef}
H_\mathrm{so}(\mathbf{x}_a,\mathbf{p}_a,\mathbf{S}_a)
= \sum_a \mathbf{\Omega}_a(\mathbf{x}_b,\mathbf{p}_b)\cdot\mathbf{S}_a,
\ee
where $\mathbf{\Omega}_a=(\Omega^i_a)$ depends on (all) the orbital degrees of
freedom $(\mathbf{x}_b,\mathbf{p}_b)$, but does not depend on the spins
$\mathbf{S}_b$. The scalar product in the Eq.\ \eqref{hsodef} is the
usual Euclidean one.

In Eq.\ \eqref{hsodef} $\mathbf{\Omega}_a$ is a priori just a notation
for the coefficient of $\mathbf{S}_a$ in $H_\mathrm{so}$. But let us now
show that it is equal to the quantity computed in the
previous section, i.e.\ the angular velocity with which the $a$th spin
vector $\mathbf{S}_a$ precesses. Indeed, the general principles of
Hamiltonian dynamics, together with the canonical Poisson brackets
\eqref{pb} and the form \eqref{hsodef}, yield
\be
\label{hamiltonspinprecess}
\dot{\mathbf{S}}_a
= \{\mathbf{S}_a,H_{\mathrm{so}}(\mathbf{x}_b,\mathbf{p}_b,\mathbf{S}_b)\}
= \mathbf{\Omega}_a(\mathbf{x}_b,\mathbf{p}_b ) \times \mathbf{S}_a.
\ee
The only difference between \eqref{hamiltonspinprecess} and the previous
result \eqref{Omega} is that, in  \eqref{hamiltonspinprecess}, $\mathbf{\Omega}_a$
is expressed in terms of canonical positions and momenta,
while in \eqref{Omega} $\mathbf{\Omega}$
it was computed in terms of (say ADM) coordinates and coordinate velocities.
Because we are working only to linear order in the spin,
and because (as was explained above)
the canonical phase-space coordinates appearing in \eqref{hsodef} and
\eqref{hamiltonspinprecess} differ from the usual ADM-type coordinates
used to express the metric [and thereby to compute the angular velocity
$\mathbf{\Omega}_a(\mathbf{x}_b^{\rm ADM},\mathbf{v}_b^{\rm ADM})$ by means of
\eqref{Vij}, \eqref{dotsic}, and \eqref{Omega}] only by terms proportional
to the spins, it suffices to use the known \cite{DJS00,DJS01a}
spinless link between ADM momenta and ADM velocities to compute
$\mathbf{\Omega}_a(\mathbf{x}_b,\mathbf{p}_b)$ from
$\mathbf{\Omega}_a(\mathbf{x}_b^{\rm ADM},\mathbf{v}_b^{\rm ADM})$.

In the previous section we introduced a specific, well-defined 
``conserved'' spin 3-vector $S_i$ to parametrize the two degrees
of freedom of a spinning particle. Our choice had the nice features of being
universally associated to the choice of a coordinate system, and of
reducing to the choice made in the flat spacetime limit \cite{BelMartin}.
However, it was by no means physically unique.

Let us now show that the freedom in the choice of  conserved spin vector
is simply a ``gauge freedom" (local rotation group) which does not
change the physical results one can deduce from the Hamiltonian.
Indeed, the condition $S_iS_i=s^2$ leaves as ambiguity in the definition of
the conserved spin variable  $S_i$ a local 3-dimensional Euclidean rotation
$ S_i \to S'_i$, with
\be
S'_i = R_{ij} S_j,
\ee
where $R$ is an arbitrary rotation matrix. It is sufficient to consider
the case of an infinitesimal rotation, say
\be
\label{rot}
R_{ij} = \delta_{ij} - \theta_{ij},
\ee
where $\theta_{ij}$ is a small antisymmetric matrix.
This leads to an infinitesimal change
\be
\delta \mathbf{S} = \btheta \times \mathbf{S},
\ee
where we introduced the dual vector $\btheta$ such
that $\theta_{ij}=\varepsilon_{ijk}\theta_k$.

Let us show that such a change can be considered as being
induced by an infinitesimal canonical transformation $g$
in the full phase space $(\mathbf{x},\mathbf{p},\mathbf{S})$.
(Canonical transformations are symmetries of Hamiltonian dynamics.
In particular they preserve the basic Poisson brackets written above.)
We recall that such a canonical transformation
acts on any phase-space function $f$ according to
\be
\delta f = \{f,g\}.
\ee
It is then easily checked that a transformation  of the form
\be
g(\mathbf{x},\mathbf{p},\mathbf{S})
= \btheta(\mathbf{x},\mathbf{p})\cdot\mathbf{S}
\ee
transforms the spin vector according to
\be
\delta \mathbf{S} = \{\mathbf{S},g\} = \btheta \times \mathbf{S},
\ee
which  exactly reproduces the effect of an infinitesimal local
rotation written above.
However, we have learned that such a local rotation must be
accompanied by a corresponding  transformation of the orbital
degrees of freedom $(\mathbf{x},\mathbf{p})$ of the form:
$\delta\mathbf{x}=\{\mathbf{x},g\}$,
$\delta\mathbf{p}=\{\mathbf{p},g\}$. Then, under the
simultaneous changes of $\mathbf{x},\mathbf{p},\mathbf{S}$ induced
by the canonical transformation $g$ (and the corresponding
change of the spin angular velocity
$\mathbf{\Omega}'\simeq\mathbf{\Omega}+\md\btheta/\md t$)
one finds that the numerical value
(evaluated at corresponding phase-space points)
of the Hamiltonian is invariant.

We have therefore shown that the arbitrariness in the ``rotational state''
of the conserved spin is simply (as expected) a ``gauge symmetry''
(under a local SO(3) group).

\section{Derivation of the spin-orbit Hamiltonian in ADM coordinates}

Let us now sketch the computation of the NLO  angular velocity $\Omega_i$
in ADM coordinates [which will then give us the NLO spin-orbit Hamiltonian
according to Eq.\ \eqref{hsodef}].

As usual we split the four-dimensional metric $g_{\mu\nu}$ into
three-dimensional objects $(\alpha,\beta_i,\gamma_{ij})$, where
\be
\label{abc}
\alpha \equiv (-g^{00})^{-1/2}, \quad
\beta_i \equiv g_{0i}, \quad
\gamma_{ij} \equiv g_{ij}.
\ee
One can show, using the definitions
$\beta^i=\gamma^{ij}\beta_j,\gamma^{ij}\gamma_{jk} =\delta_{ik}$,
that the following exact formulas hold:
\begin{subequations}
\label{Chrsym}
\begin{align}
{\Gamma^0}_{0i}
&= \frac{1}{\alpha} \Big(\alpha_{,i}+K_{ij}\beta^j\Big),
\\[1ex]
{\Gamma^0}_{ij} &= \frac{1}{\alpha} K_{ij},
\\[1ex]
{\Gamma^i}_{j0} &= \frac{1}{2}\gamma^{ik}\gamma_{kj,0}
- \frac{1}{\alpha} \beta^i \alpha_{,j}
+ \frac{1}{2}\gamma^{ik} (\beta_{k,j}-\beta_{j,k})
-\frac{1}{\alpha}\beta^i\beta^k K_{kj},
\\[1ex]
{\Gamma^i}_{jk}
&= {^3}\Gamma^i_{jk} - \frac{1}{\alpha} \beta^i K_{jk},
\end{align}
\end{subequations}
where $K_{ij}$ is the extrinsic curvature of the constant time slice.
Note that, for convenience, we use the  $K_{ij}\sim +\;\dot{\gamma_{ij}}$
sign convention (instead of the $-\;\dot{\gamma_{ij}}$ convention used
e.g.\ in Ref.\ \cite{MTW73}). In terms of the field momenta $\pi^{ij}$
it reads,
\be
\label{kij}
K_{ij} = \frac{16\pi G}{c^3}
\frac{1}{\sqrt{\gamma}} \Big(\gamma_{ik}\gamma_{jl}
-\frac{1}{2}\gamma_{ij}\gamma_{kl}\Big)\pi^{kl},
\ee
where $\gamma=\det(\gamma_{ij})$. The Christoffel symbols related
with the 3-metric $\gamma_{ij}$ are denoted by ${^3}\Gamma^i_{jk}$.
Let us also note that the dimensionless coordinate velocity $v^i$
can be expressed in terms of the {\it bare kinematical} linear momenta 
$p^{\rm bare}_i= m c u_i$, in full
generality, as follows
\be
v^i = \frac{\alpha\gamma^{ij}p^{\rm bare}_j}
{(m^2c^2+\gamma^{kl}p^{\rm bare}_k p^{\rm bare}_l)^{1/2}}
- \gamma^{ij}\beta_j.
\ee
Note, however, that the latter result applies to the {\it canonical}
momentum only modulo corrections proportional to the spin.

We employ the ADMTT (ADM transverse-traceless)
coordinate conditions \cite{ADM62}
\be
\label{admtt}
\gamma_{ij} = \bigg(1+\frac{1}{8}\phi\bigg)^4 \delta_{ij}
+ h^{\mathrm{TT}}_{ij}, \quad \pi^{ii} = 0,
\ee
and recall that
\be
\label{splitoffm}
\pi^{ij} = \tilde{\pi}^{ij} + \pi_\mathrm{TT}^{ij}
\ee
with $\pi_\mathrm{TT}^{ij}$ being of the order $1/c^5$ \cite{JS98}.

Let us now expand all quantities in a post-Newtonian (PN) expansion.
Here and below the subscript $(n)$ indicates the part of a quantity which
is of the $n$th post-Newtonian order, i.e.\ which is proportional to $(1/c^2)^n$.
For instance we decompose:
\be
\label{omega2&omega4}
\Omega_i = \Omega_{(2)i} + \Omega_{(4)i} + {\cal O}(c^{-6}).
\ee
Here  $\Omega_{(2)i} \propto G/c^2$ is the well-known LO contribution
\cite{Barker:1975ae,DamourVarenna75,BornerEhlersRudolph,D'Eath:1975vw},
while $\Omega_{(4)i} \propto G/c^4 + G^2/c^4$
is the NLO contribution that we wish to compute.
These contributions are more explicitly given
in terms of the ``precession velocity'' $\tilde{V}^{ij}$
of the ``coordinate spin vector'' $\tilde{S}_i$, which entered Eq.\ \eqref{dotsi}.
Inserting in Eq.\ \eqref{Vij} the Christoffel symbols \eqref{Chrsym},
and then inserting the result in Eq.\ \eqref{dotsic}
[where $H^{ij}$ is computed from Eqs.\ \eqref{Gdef}, \eqref{GHH},
\eqref{abc}, \eqref{kij}, \eqref{admtt}, and \eqref{splitoffm}],
we obtain the following more explicit formulas
for the 3-vectors $\Omega_{(2)i}$ and $\Omega_{(4)i}$
from Eq.\ \eqref{omega2&omega4}:
\begin{subequations}
\label{omega3&omega5}
\begin{align}
\label{omega3}
\Omega_{(2)i}/c &= \frac{1}{2} \varepsilon_{ijk} \bigg(
\beta_{(3)j,k}+\Big(\alpha_{(2),j}-\frac{1}{2}\phi_{(2),j}\Big)v^k \bigg),
\\[2ex]
\label{omega5}
\Omega_{(4)i}/c &= \frac{1}{2} \varepsilon_{ijk} \bigg(
\beta_{(5)j,k}
+ \beta_{(3)k}\alpha_{(2),j}
- \frac{1}{2}\phi_{(2)}\beta_{(3)j,k}
+ \frac{1}{16}\phi_{(2)}\phi_{(2),j}v^k
- \frac{1}{2}\phi_{(4),j}v^k
- h^{\mathrm{TT}}_{(4)kl,j}v^l
\nonumber\\[1ex]&\quad
+ (\alpha_{(4),j}-\alpha_{(2)}\alpha_{(2),j})v^k
+ \tilde{\pi}_{(3)}^{jl}v^kv^l
- \frac{1}{2}\alpha_{(2),k}v^jv^lv^l
+ \frac{1}{4}\frac{\dot{v}^j}{c}v^kv^lv^l \bigg).
\end{align}
\end{subequations}

At this point, it only remains to implement three technical steps: 
(i) to insert the explicit form of the 2PN-accurate
metric describing two spin-less particles in ADMTT coordinates 
(from \cite{OOKH74} and \cite{JS98}),
(ii) to replace the velocities $v^i$ by their 1PN-accurate expression
in terms of the canonical momenta $p_i$, and, finally,
(iii) to regularize the self-interaction terms
that arise when evaluating Eqs.\ \eqref{omega3&omega5}.

The explicit expressions for the metric functions entering Eqs.\ \eqref{omega3&omega5}
can be found e.g.\ in Appendix~A of Ref.\ \cite{JS98}
(where the functions $\phi_{(2)}$, $\phi_{(4)}$,
$\tilde{\pi}_{(3)}^{ij}$, and $h^\mathrm{TT}_{(4)ij}$ can be found)
and in Ref.\ \cite{OOKH74} (where the functions $\alpha_{(2)}$, $\alpha_{(4)}$
and $\beta_{(3)i}$, $\beta_{(5)i}$ are given).

As for reexpressing the velocities in terms of momenta, it yields a further
PN-expansion of the form $v_a^i=v_{a(1)}^i+v_{a(3)}^i+O(1/c^5)$,
where\footnote{Here and below $a,b=1,2$ are the particles' labels,
so $m_a$, $\mathbf{x}_a=(x_a^i)$, and $\mathbf{p}_a=(p_{ai})$
denote, respectively, the mass parameter, the position vector,
and the linear momentum vector of the $a$th body; for $a\ne b$ we also define
$\mathbf{r}_{ab}\equiv\mathbf{x}_a-\mathbf{x}_b$,
$r_{ab}\equiv|\mathbf{r}_{ab}|$, $\mathbf{n}_{ab}\equiv\mathbf{r}_{ab}/r_{ab}$;
$|\cdot|$ stands here for the Euclidean length of a 3-vector.}
$v_{a(1)}^i$ is the coordinate velocity of the $a$th
particle expressed in terms of the canonical variables $\mathbf{x}_a$ and
$\mathbf{p}_a$ at the Newtonian accuracy, i.e., $v_{a(1)}^i={p_{ai}/(m_ac)}$,
and $v_{a(3)}^i$ is the 1PN correction to ${p_{ai}/(m_ac)}$.
The latter 1PN correction explicitly reads
\be
\label{v1pn}
v_{1(3)}^i = \frac{G\npii}{2c^3r_{12}}\,n^i_{12}
+ \Big(-\frac{\pipi}{2m_1^3c^3}-\frac{3Gm_2}{m_1c^3r_{12}}\Big)\,p_{1i}
+ \frac{7G}{2c^3r_{12}}\,p_{2i},
\ee
the expression for $v_{2(3)}^i$ can be obtained from the above by exchanging the
particles' labels.

The final step then consists in evaluating (note that the meaning of
$\Omega^i_{a(2)}$ and $\Omega^i_{a(4)}$ is now slightly different because of
the re-expansion of velocities in a PN expansion)
\begin{subequations}
\label{omega}
\begin{align}
\label{omega3a}
\Omega^i_{a(2)}/c &\equiv \frac{1}{2} \varepsilon_{ijk}
{\rm Reg}_a \bigg( \beta_{(3)j,k}
+ \Big(\alpha_{(2),j}-\frac{1}{2}\phi_{(2),j}\Big)v_{a(1)}^k \bigg),
\\[2ex]
\label{omega5a}
\Omega^i_{a(4)}/c &\equiv \frac{1}{2} \varepsilon_{ijk}
 {\rm Reg}_a \bigg( \beta_{(5)j,k} + \beta_{(3)k}\alpha_{(2),j}
-\frac{1}{2}\phi_{(2)}\beta_{(3)j,k}
+ \frac{1}{16}\phi_{(2)}\phi_{(2),j}v_{a(1)}^k
- \frac{1}{2}\phi_{(4),j}v_{a(1)}^k
- h^{\mathrm{TT}}_{(4)kl,j}v_{a(1)}^l
\nonumber\\[1ex]&\quad
+ (\alpha_{(4),j}-\alpha_{(2)}\alpha_{(2),j})v_{a(1)}^k
+ \tilde{\pi}_{(3)}^{jl}v_{a(1)}^kv_{a(1)}^l
- \frac{1}{2}\alpha_{(2),k}v_{a(1)}^jv_{a(1)}^lv_{a(1)}^l
+ \frac{1}{4} \frac{\dot{v}_{a(1)}^j}{c}v_{a(1)}^kv_{a(1)}^lv_{a(1)}^l
\nonumber\\[1ex]&\quad
+ \Big(\alpha_{(2),j}-\frac{1}{2}\phi_{(2),j}\Big)v_{a(3)}^k
\bigg),
\end{align}
\end{subequations}
where ${\rm Reg}_a\big(f(\mathbf{x})\big)$ indicates that one must regularize
the limit $\mathbf{x}\to\mathbf{x}_a$. At the level at which we are working,
this regularization is not ambiguous and can, for instance, be simply performed
by using Hadamard's ``partie finie'' regularization
(as explained, e.g., in Appendix~B of Ref.\ \cite{JS98}).
The final results we got read
\begin{subequations}
\label{omegafinal}
\begin{align}
\mathbf{\Omega}_{1(2)} &= \frac{G}{c^2r_{12}^2}
\bigg( \frac{3m_2}{2m_1}\ncpi - 2\ncpii \bigg),
\\[2ex]
\mathbf{\Omega}_{1(4)} &= \frac{G^2}{c^4r_{12}^3} \Bigg(
\bigg(-\frac{11}{2}m_2-5\frac{m_2^2}{m_1}\bigg)\ncpi
+ \bigg(6m_1+\frac{15}{2}m_2\bigg)\ncpii \Bigg)
\nonumber\\[1ex]&\quad
+ \frac{G}{c^4r_{12}^2} \Bigg( \bigg(
- \frac{5m_2\pipi}{8m_1^3} - \frac{3\pipii}{4m_1^2}
+ \frac{3\piipii}{4m_1m_2}
- \frac{3\npi\npii}{4m_1^2} - \frac{3\npii^2}{2m_1m_2} \bigg)\ncpi
\nonumber\\[1ex]&\quad
+ \bigg(\frac{\pipii}{m_1m_2}+\frac{3\npi\npii}{m_1m_2}\bigg)\ncpii
+ \bigg( \frac{3\npi}{4m_1^2} - \frac{2\npii}{m_1m_2} \bigg)\cpipii
\Bigg).
\end{align}
\end{subequations}
The expressions for $\mathbf{\Omega}_{2(2)}$ and $\mathbf{\Omega}_{2(4)}$
can be obtained from the above formulas by exchanging the particles' labels.

{}From these results we can then explicitly write the spin-orbit
Hamiltonian to leading and next-to-leading PN orders. Indeed,
\be
\label{hsodefnew}
H_\mathrm{so}(\mathbf{x}_a,\mathbf{p}_a,\mathbf{S}_a)
= \sum_a \mathbf{\Omega}_a(\mathbf{x}_b,\mathbf{p}_b)\cdot\mathbf{S}_a
= \sum_a \big({\bf\Omega}_{a(2)}({\bf x}_b,{\bf p}_b)
+ {\bf\Omega}_{a(4)}({\bf x}_b,{\bf p}_b)\big)\cdot\mathbf{S}_a.
\ee
More explicitly, the separate LO and NLO contributions
in the PN expansion of the spin-orbit interaction term,
\be
\label{hsot}
H_{\mathrm{so}}({\bf x}_a,{\bf p}_a,{\bf S}_a)
= \frac{1}{c^2}\,H^\mathrm{LO}_\mathrm{so}({\bf x}_a,{\bf p}_a,{\bf S}_a)
+ \frac{1}{c^4}\,H^\mathrm{NLO}_\mathrm{so}({\bf x}_a,{\bf p}_a,{\bf S}_a)
+ {\cal O}\left(\frac{1}{c^6}\right),
\ee
read,
\begin{subequations}
\label{hso}
\begin{align}
H^\mathrm{LO}_\mathrm{so}({\bf x}_a,{\bf p}_a,{\bf S}_a)
&= c^2 \sum_a {\bf\Omega}_{a(2)}({\bf x}_b,{\bf p}_b)\cdot{\bf S}_a,
\\
H^\mathrm{NLO}_\mathrm{so}({\bf x}_a,{\bf p}_a,{\bf S}_a)
&= c^4 \sum_a {\bf\Omega}_{a(4)}({\bf x}_b,{\bf p}_b)\cdot{\bf S}_a.
\end{align}
\end{subequations}

Finally, note a remarkable feature of our Hamiltonian approach to spin-orbit
effects:  the sole computation of the {\it rotational} velocity of the (``conserved'')
spin vector (given by parallel transport in the 2PN-accurate metric of $N$ {\it spinless} bodies)
determines the NLO {\it spin-dependent} terms in the {\it translational} equations of motion of
$N$ {\it spinning} particles. Indeed, the sole knowledge of
$\mathbf{\Omega}_a(\mathbf{x}_b,\mathbf{p}_b)$ yields that of the total 
spin-dependent Hamiltonian \eqref{hcomplete} with \eqref{hsodef}, so that the
general principles of Hamiltonian dynamics (with canonical Poisson brackets) yield
\be
\label{translational}
\dot{\mathbf{x}}_a = + \frac{\partial H(\xpsb)}{\partial\mathbf{p}_a},
\quad
\dot{\mathbf{p}}_a = - \frac{\partial H(\xpsb)}{\partial\mathbf{x}_a}.
\ee
In view of the availability of algebraic manipulation programmes, there is no
need to write down explicitly the translational equations of motion \eqref{translational},
with NLO accuracy in spin-orbit terms (and 3PN accuracy in spin-independent terms 
\cite{DJS00,DJS01a}). We shall verify below that the Hamiltonian, ADM-coordinate
translational equations of motion \eqref{translational} are equivalent to the
harmonic-coordinate ones recently derived in \cite{FBB06,BBF06} by a more complex
calculation which involved the explicit consideration of spin-dependent contributions
in the metric.

\section{Poincar\'e invariance}

The general relativistic dynamics of an {\it isolated} $N$-body system
should admit the full Poincar\'e group as a {\it global} symmetry
(because it is a symmetry which preserves asymptotic flatness).
On the other hand, this symmetry is {\it not manifest} in the Hamiltonian
ADM approach to the $N$-body dynamics because it splits space and time,
and uses  non-Lorentz covariant coordinate conditions.
In a previous paper \cite{DJS00}, treating non-spinning particles,
the authors showed how to bypass this technical mismatch: the basic idea
is that, in the Hamiltonian formalism, the global Poincar\'e symmetry is
realized in phase-space in a non-linear manner. However, one can efficiently
detect the presence of this symmetry by proving the existence of ten phase-space
generators $H(\xpsa)$, $P_i(\xpsa)$, $J_i(\xpsa)$, $G_i(\xpsa)$
(depending on all phase-space variables) whose Poisson brackets reproduce
the standard Poincar\'e algebra. In the case of non-spinning particles,
Ref.\ \cite{DJS00} constructed  the ten generators of the Poincar\'e group at
the 3PN level of approximation. We shall show here how to extend this construction
to the more involved case of a system of spinning particles.

Let us first recall the explicit Poisson-bracket form
of the Poincar\'e algebra that should be realized:
\begin{subequations}
\begin{align}
\label{PB1}
\{P_i,P_j\} &= 0, \quad \{P_i,H\} = 0, \quad \{J_i,H\} = 0,
\\[2ex]
\label{PB2}
\{J_i,P_j\} &= \varepsilon_{ijk}\,P_k,
\quad \{J_i,J_j\} = \varepsilon_{ijk}\,J_k,
\\[2ex]
\label{PB3}
\{J_i,G_j\} &= \varepsilon_{ijk}\,G_k,
\\[2ex]
\label{PB4}
\{G_i,H\} &= P_i,
\\[2ex]
\label{PB5}
\{G_i,P_j\} &= \frac{1}{c^2}\,H\,\delta_{ij},
\\[2ex]
\label{PB6}
\{G_i,G_j\} &= -\frac{1}{c^2}\,\varepsilon_{ijk}\,J_k.
\end{align}
\end{subequations}

The translation, $P_i$, and rotation, $J_i$, generators are simply realized as
\begin{subequations}
\begin{align}
P_i(\xpsa) &= \sum_a p_{ai},
\\[1ex]
J_i(\xpsa) &= \sum_a
\big(\varepsilon_{ik\ell}\,x_a^k\,p_{a\ell} + S_{ai}\big).
\end{align}
\end{subequations}
Note the very simple, additive,  form of these generators, and,
in particular, how our Hamiltonian ``conserved spin'' variables
appear as Newtonian-like (but relativistically correct) contributions.

As for the Hamiltonian $H$, we already know that (in our linear-in-spin approximation),
it is a sum of an orbital part, $H_\mathrm{o}$,
and of the above-determined spin-orbit part, $H_\mathrm{so}$,
Eqs.\ \eqref{hsodefnew}, \eqref{hsot}, and \eqref{hso}:
\be
H({\bf x}_a,{\bf p}_a,{\bf S}_a)
= H_{\mathrm{o}}({\bf x}_a,{\bf p}_a)
+ H_{\mathrm{so}}({\bf x}_a,{\bf p}_a,{\bf S}_a).
\ee
The orbital Hamiltonian $H_{\mathrm{o}}$ (including the rest-mass
contribution) is explicitly known up to the 3PN order
\cite{DJS00,DJS01a}:
\be
H_{\mathrm{o}}({\bf x}_a,{\bf p}_a) = \sum_a m_a c^2
+ H_{\text{oN}}({\bf x}_a,{\bf p}_a)
+ \frac{1}{c^2}\,H_{\rm o1PN}({\bf x}_a,{\bf p}_a)
+ \frac{1}{c^4}\,H_{\rm o2PN}({\bf x}_a,{\bf p}_a)
+ \frac{1}{c^6}\,H_{\rm o3PN}({\bf x}_a,{\bf p}_a)
+ {\cal O} \left( \frac{1}{c^8} \right).
\ee

The most delicate generator to consider is the boost (or center-of-mass)
vector ${\mathbf{G}}$. It can be represented
as a sum of ``orbital'' and ``spin-orbit'' parts
\be
\label{gtot}
{\mathbf{G}}({\bf x}_a,{\bf p}_a,{\bf S}_a)
= {\mathbf{G}}_{\mathrm{o}}({\bf x}_a,{\bf p}_a)
+ {\mathbf{G}}_{\mathrm{so}}({\bf x}_a,{\bf p}_a,{\bf S}_a),
\ee
where, as everywhere in this paper, we call ``spin-orbit'' the part
which is linear in the spin variables.
The orbital part, ${\mathbf{G}}_{\mathrm{o}}$, was explicitly determined up
to the the 3PN order in Ref.\ \cite{DJS00}:
\be
{\mathbf{G}}_{\mathrm{o}}({\bf x}_a,{\bf p}_a)
= \sum_a m_a {\mathbf{x}}_a
+ \frac{1}{c^2}\,{\mathbf{G}}_{\rm o1PN}({\bf x}_a,{\bf p}_a)
+ \frac{1}{c^4}\,{\mathbf{G}}_{\rm o2PN}({\bf x}_a,{\bf p}_a)
+ \frac{1}{c^6}\,{\mathbf{G}}_{\rm o3PN}({\bf x}_a,{\bf p}_a)
+ {\cal O} \left( \frac{1}{c^8} \right).
\ee
The spin-orbit part can be decomposed in leading-order (LO),
next-to-leading-order (NLO), and further contributions:
\be
\label{gsopn}
\mathbf{G}_\mathrm{so}({\bf x}_a,{\bf p}_a,{\bf S}_a)
= \frac{1}{c^2}\,
\mathbf{G}^\mathrm{LO}_\mathrm{so}({\bf x}_a,{\bf p}_a,{\bf S}_a)
+ \frac{1}{c^4}\,
\mathbf{G}^\mathrm{NLO}_\mathrm{so}({\bf x}_a,{\bf p}_a,{\bf S}_a)
+ {\cal O} \left( \frac{1}{c^6} \right).
\ee
The leading-order term in \eqref{gsopn} is known from the special-relativistic
limit (by replacing the special-relativistic energy in the results of, e.g.,
Refs.\ \cite{BelMartin,HansonRegge}, by the rest-mass contribution)
\be
\mathbf{G}^\mathrm{LO}_\mathrm{so}(\mathbf{x}_a,\mathbf{p}_a,\mathbf{S}_a)
= -\frac{\csipi}{2m_1} + (1\leftrightarrow2),
\ee
where the operation ``$+(1\leftrightarrow2)$'' denotes the addition
to each displayed term of another one obtained by exchanging the
particles' labels.

The real difficulty lies in constructing the NLO contribution to the
boost generator (and in proving that it satisfies the correct Poincar\'e
algebra displayed above). We solved this problem by using (as in our
previous work \cite{DJS00}) the method of undetermined coefficients.
The most general form of $\mathbf{G}^\mathrm{NLO}_\mathrm{so}$
can a priori depend on eight unknown dimensionless numerical coefficients
$g_1,\ldots,g_8$:
\begin{align}
\label{gso2pngs}
\mathbf{G}^\mathrm{NLO}_\mathrm{so} &= \frac{\pipi}{8m_1^3} \csipi
+ \frac{Gm_2}{r_{12}} \Bigg(
g_1\frac{\csipi}{m_1} +  g_2\frac{\csipii}{m_2}
+ \bigg(g_3\frac{\npi}{m_1}+g_4\frac{\npii}{m_2}\bigg) \cnsi
\nonumber\\[1ex]&\quad
+ \Big(g_5\frac{\sinpi}{m_1}+g_6\frac{\sinpii}{m_2}\Big)
{\mathbf{n}}_{12}
\Bigg) + \frac{Gm_2}{r_{12}^2} \Bigg(g_7 \frac{\sinpi}{m_1}
+ g_8 \frac{\sinpii}{m_2} \Bigg) {\mathbf{x}}_1
+ (1\leftrightarrow2),
\end{align}
where we have introduced the following notation for the Euclidean
mixed product of 3-vectors: $(V_1,V_2,V_3)\equiv
\mathbf{V}_1\cdot(\mathbf{V}_2\times\mathbf{V}_3)
=\varepsilon_{ijk}V_1^iV_2^jV_3^k$. Note that the coefficient of the first term
on the right-hand-side is determined by considering the special-relativistic
limit \cite{Souriau,BelMartin,HansonRegge}. We have also used some structural information
coming from a conceivable field-theory computation of $\mathbf{G}$ (say as the 
space integral of the $0i$ component of some effective stress-energy tensor).
Indeed, such a computation could be thought of in terms of some Feynman-like
diagrams, where the interaction terms (i.e.\ those containing a power of $G$)
would all be proportional to some basic ``source'' term involving
either $S_1$ and $m_2$ (connected by a propagator, and possibly some power
of the velocities, $v_1 \sim p_1/m_1$ or $v_2 \sim p_2/m_2$), or similar
terms involving $S_2$ and $m_1$. The main point being that pure ``self-interaction''
terms (say proportional to $S_1$ and $m_1$) cannot appear.

Let us now consider the explicit Poincar\'e algebra requirements of
Eqs.\ \eqref{PB1}--\eqref{PB6}. It is easily verified
that the generators $P_i$, $J_i$, $H$, and $G_i$, in the forms given above,
exactly satisfy the relations \eqref{PB1}, \eqref{PB2}, and \eqref{PB3}.
We now consider whether the center-of-mass vector $\mathbf{G}$
with the 2PN spin-orbit part given by Eq.\ \eqref{gso2pngs} can
satisfy the three relations \eqref{PB4}--\eqref{PB6}. This requirement yields
many equations that have to be satisfied by the unknown coefficients 
$g_1,\ldots,g_8$. We have first found that there exist {\em unique}
values of the coefficients $g_1,\ldots,g_8$ ensuring the fulfillment
of the sole relation \eqref{PB4}. These values are
\begin{align}
\label{gsol}
g_1 = \frac{5}{4}, \;
g_2 = -\frac{3}{2}, \;
g_3 = 0, \;
g_4 = -\frac{1}{2}, \;
g_5 = -\frac{1}{4}, \;
g_6 = 1, \;
g_7 = \frac{3}{2}, \;
g_8 = -2.
\end{align}
Then we have checked that the solution \eqref{gsol} also guarantees
the fulfillment of the remaining relations \eqref{PB5} and \eqref{PB6}.

In summary, we succeeded in proving the Poincar\'e invariance
of the above-defined NLO spin-orbit interaction
[determined by Eqs.\ \eqref{hsodefnew}, \eqref{hsot}, and \eqref{hso}]
by explicitly constructing ten phase-space generators
satisfying the Poincar\'e algebra brackets of Eqs.\ \eqref{PB1}--\eqref{PB6}.

\section{Comparison with harmonic-coordinate-based results} 

References \cite{FBB06,BBF06} recently computed, by means of two separate
calculations and in {\it harmonic coordinates}, both the NLO spin-dependent
contributions in the translational equations of motion of two spinning particles,
and the corresponding NLO terms in the spin precessional equations of motion.
In the present Section we shall prove that our results are physically
equivalent to the results of Refs.\ \cite{FBB06,BBF06} by finding
the explicit form of the transformation that match the ADM variables
used by us with the harmonic variables used in Refs.\ \cite{FBB06,BBF06}.
Let us start by warning the reader that in the whole paper \cite{FBB06} and
in  most of the paper \cite{BBF06} Blanchet et al.\ chose to express
their results in terms of some ``non-conserved'' spin variables $\sbbfnca$,
i.e.\ variables whose Euclidean magnitudes are not conserved in time.
It is only in Sec.\ VII of \cite{BBF06} that redefined spin variables with conserved
Euclidean lengths, say $\sbbfa$, are introduced and used.

Our task here will be to exhibit the explicit transformation
between the ``ADM variables'' $(\xpsa)$ used in our work,
and the ``harmonic variables''
$(\mathbf{y}_a,\mathbf{v}_a\equiv\dot{\mathbf{y}}_a,\sbbfa)$ used in \cite{FBB06,BBF06},
and to prove that this transformation maps the two sets of results into each other.
(It is more convenient for us to exhibit the link with the ``conserved'' version of
the harmonic spin variable used by Blanchet et al. The relation between their
two spin variables, $\sbbfnca$ and $\sbbfa$, is given in Eq.\ (7.4) of \cite{BBF06}.)

We write the transformation of variables in the general form\footnote{
Here, both sides refer to the same numerical value of their respective
coordinate times.}
\begin{subequations}
\label{transformations}
\begin{align}
\mathbf{y}_a(t)
&= \mathbf{Y}_a(\mathbf{x}_b(t),\mathbf{p}_b(t),\mathbf{S}_b(t)),
\\[1ex]
\sbbfa(t)
&= {\bf\Sigma}_a(\mathbf{x}_b(t),\mathbf{p}_b(t),\mathbf{S}_b(t)).
\end{align}
\end{subequations}
Let us first find the transformation ${\bf\Sigma}_a$ between spin variables.
Section VII of Ref.\ \cite{BBF06} gives (see Eq.\ (7.6) there) the explicit result for
the angular velocity vector ${\mathbf{\Omega}}_a^{\mathrm{BBF}}$ of their
conserved harmonic spin variable, yielding a spin precessional equation of
motion of the form
\be
\frac{\md\sbbfa}{\md t} = {\bf\Omega}^\mathrm{BBF}_a\times\sbbfa,
\quad a=1,2.
\ee
They give the NLO expression of ${\mathbf{\Omega}}_a^{\mathrm{BBF}}$ in terms of
the harmonic orbital coordinates $(\mathbf{y}_a,\mathbf{v}_a)$. We
have re-expressed ${\mathbf{\Omega}}_a^{\mathrm{BBF}}(\mathbf{y}_b,\mathbf{v}_b)$ 
in terms of ADM
coordinates and momenta, to 1PN accuracy (using the well-known link between the two
sets of variables\footnote{We recall that harmonic and ADM {\it coordinates} coincide
at 1PN, but that one must transform velocities into momenta by means of the
1PN transformation Eq.\ \eqref{v1pn}. See \cite{DJS01b} for the 3PN-accurate version
of this transformation.}). We then compared the result with our results \eqref{omegafinal}. We
have found
\begin{subequations}
\label{omegaha}
\begin{align}
{\bf\Omega}_{a(2)}^\mathrm{BBF}(\mathbf{y}_b,\mathbf{v}_b)
&= {\bf\Omega}_{a(2)}(\mathbf{x}_b,\mathbf{p}_b),
\\[1ex]
{\bf\Omega}_{a(4)}^\mathrm{BBF}(\mathbf{y}_b,\mathbf{v}_b)
&= {\bf\Omega}_{a(4)}(\mathbf{x}_b,\mathbf{p}_b)
+ \frac{\md\btheta_a}{\md t},
\end{align}
\end{subequations}
where
\be
\label{theta}
\btheta_1 = \frac{G}{c^4r_{12}} \bigg( - \frac{\npii}{4m_1}\ncpi
+ \frac{\npii}{m_2}\ncpii - \frac{9}{4m_1}\cpipii \bigg).
\ee
{}From the results \eqref{omegaha}--\eqref{theta} it is easy to deduce
that the two sets of spin precession equations of motion are physically
equivalent\footnote{As a further check, we have also explicitly verified
that the Hamiltonian time derivative (computed with our dynamics, namely $\{\sbbfnca,H\}$)
of the originally defined (non-conserved) spin vector $\sbbfnca$ of \cite{FBB06}
coincides with the NLO spin precession law given by Eqs.\ (6.1)--(6.3) there. To do this
calculation we defined the phase-space quantity $\sbbfnca(\xpsb)$ by inserting
Eqs.\ \eqref{omegaha}, \eqref{theta} into Eq.\ (7.6) of \cite{BBF06}.},
and that the two sets of spin variables are related as in the
general transformation links written above with
a spin transformation $\mathbf\Sigma_a$ of the explicit form:
\be
\label{Sigmaexpl}
\mathbf\Sigma_a(\mathbf{x}_b,\mathbf{p}_b,\mathbf{S}_b)
= \mathbf{S}_a
+ \btheta_a(\mathbf{x}_b,\mathbf{p}_b)\times\mathbf{S}_a.
\ee
In other words, our conserved spin variable differs from the conserved
spin variable defined in Eq.\ (7.4) of \cite{BBF06} by a small (time-dependent)
rotation of angle $\btheta_a(\mathbf{x}_b,\mathbf{p}_b)$. Such a
difference was a priori to be expected because constant-magnitude spin vectors
are not uniquely defined. We have shown above that to each choice of
coordinate system is canonically associated a particular choice of
local orthonormal frame (along the worldline of a spinning particle),
and thereby a particular choice of ``conserved'' spin 3-vector. We have
investigated whether the  conserved spin 3-vector
defined by Blanchet et al.\ does correspond to applying our general definition
to the case of  harmonic coordinates. The answer is ``no''. We found that
if Blanchet et al.\ had used our general definition \eqref{sic}
in their harmonic coordinate system, the angular velocity ${\bf\Omega}_a$
that they would have obtained would differ from our ADM spin vector by a 
rotation vector ${\btheta}_a$ differing from the result above by having
the factor 9 replaced by 1 in the last term of Eq.\ \eqref{theta}. There is nothing
surprising in such a difference as the spin re-definition used by
Blanchet et al.\ was somewhat arbitrary. Anyway, as already mentioned above
physical results will not depend on such ``gauge choices.''

Let us now turn to the determination of the transformation ${\bf{Y}}_a$ 
between ADM and harmonic {\it orbital} degrees of freedom. As usual we
can decompose ${\bf{Y}}_a$ into spin-independent,
${\bf{Y}}^{\mathrm{o}}_{a}$, and spin-dependent (and linear-in-spin),
${\bf{Y}}^{\mathrm{so}}_{a}$, terms:
\be
{\bf{Y}}_a({\mathbf{x}}_b,{\mathbf{p}}_b,{\mathbf{S}}_b)
= {\mathbf{x}}_a
+ {\bf{Y}}^{\mathrm{o}}_{a}({\mathbf{x}}_b,{\mathbf{p}}_b)
+ {\bf{Y}}^{\mathrm{so}}_{a}
({\mathbf{x}}_b,{\mathbf{p}}_b,{\mathbf{S}}_b),
\ee
where the spin-dependent term is of the form
\be
{\bf{Y}}^{\mathrm{so}}_a({\mathbf{x}}_b,{\mathbf{p}}_b,{\mathbf{S}}_b)
=  {\bf{Y}}^{\mathrm{so}}_{a(2)}
({\mathbf{x}}_b,{\mathbf{p}}_b,{\mathbf{S}}_b)
+ {\bf{Y}}^{\mathrm{so}}_{a(4)}
({\mathbf{x}}_b,{\mathbf{p}}_b,{\mathbf{S}}_b)
+ {\cal O}(c^{-6}).
\ee

The spin-independent part of the transformation was explicitly given,
up to the 3PN order, in Ref.\ \cite{DJS01b}.
The leading order spin-dependent part has been known for
many years (see, e.g., Ref.\ \cite{DS88}), and equals
\be
{\bf{Y}}^{\mathrm{so}}_{a(2)}
({\mathbf{x}}_b,{\mathbf{p}}_b,{\mathbf{S}}_b)
= \frac{\sapav}{2m_a^2c^2}.
\ee
We have determined the next-to-leading order spin-dependent part,
$\mathbf{Y}^\mathrm{so}_{a(4)}$, by using again the method of undetermined
coefficients. We have considered the most general template for
$\mathbf{Y}^\mathrm{so}_{a(4)}$ which depends (after using the special relativistic
limit to determine the $1/c^4$ term which remains in the $G\to0$ limit,
and structural information of the same type as that explained above in
the case of $\mathbf{G}_\mathrm{so}$)\footnote{More specifically we required
that, say, $m_1 \mathbf{Y}^\mathrm{so}_{1}$ be proportional (modulo
some velocity-dependent factors involving $v_a\sim p_a/m_a$) either to $m_1 S_2$
or to $m_2 S_1$.} on 12 unknown coefficients. It reads
\begin{align}
{\bf{Y}}^{\mathrm{so}}_{1(4)}
({\mathbf{x}}_a,{\mathbf{p}}_a,\mathbf{S}_a)
&= -\frac{\pipi}{8 c^4 m_1^4} \sipiv
+ \frac{Gm_2}{c^4r_{12}}\frac{1}{m_1}
\Bigg( a_1 \frac{\sipiv}{m_1} + a_2 \frac{\sipiiv}{m_2} 
\nonumber\\&\quad
+ \bigg( a_3 \frac{\npi}{m_1} + a_4 \frac{\npii}{m_2} \bigg) \nsi
+ \bigg( a_5 \frac{\sinpi}{m_1} + a_6 \frac{\sinpii}{m_2}\bigg)
{\bf{n}}_{12}
\Bigg)
\nonumber\\&\quad
+ \frac{G}{c^4r_{12}} \Bigg( b_1 \frac{\siipiv}{m_1}
+ b_2 \frac{\siipiiv}{m_2}
+ \bigg( b_3 \frac{\npi}{m_1} + b_4 \frac{\npii}{m_2} \bigg) \nsii
\nonumber\\&\quad
+ \bigg( b_5 \frac{\siinpi}{m_1} + b_6 \frac{\siinpii}{m_2}\bigg)
{\bf{n}}_{12} \Bigg).
\end{align}
One can now think of two different ways of determining whether there exists
a set of coefficients $a_1,\ldots,a_6;b_1,\ldots,b_6$ such that our
translational Hamiltonian equations of motion (with NLO spin-dependent terms),
Eq.\ \eqref{translational}, are physically equivalent to the corresponding
translational harmonic equations of motion derived in \cite{FBB06}.
(1) A first way would consist of inserting the putative general transformation
${\bf{Y}}_a(a_1,\ldots,a_6 ; b_1,\ldots,b_6)$ directly into the translational
equations of motion derived in \cite{FBB06} (using the fact that we have already
determined how their spin variables are linked to ours), and to compare the result to
the explicit form of our translational Hamiltonian equations of motion,
Eq. \eqref{translational}. This approach is, however, computationally heavy.
(2) Therefore, we have instead used a simpler approach consisting in comparing
the ten conserved quantities derived in harmonic coordinates in Ref.\ \cite{FBB06},
namely the energy $E(\mathbf{y}_a,\mathbf{v}_a,\sbbfnca)$,
the total linear momentum ${\mathbf{P}}({\mathbf{y}}_a,{\mathbf{v}}_a,\sbbfnca)$,
the total angular momentum ${\mathbf{J}}({\mathbf{y}}_a,{\mathbf{v}}_a,\sbbfnca)$,
and the center-of-mass vector $\mathbf{G}(\mathbf{y}_a,\mathbf{v}_a,\sbbfnca)$,
with the ten phase-space Poincar\'e generators constructed above within our
Hamiltonian formalism. To do this comparison explicitly, we first need to perform
two replacements: (i) to replace the non-conserved spin variable $\sbbfnca$ used 
in \cite{FBB06} in terms of the conserved one $\sbbfa$ introduced in \cite{BBF06},
thereby obtaining new expressions $E(\mathbf{y}_a,\mathbf{v}_a,\sbbfa)$,
${\mathbf{P}}({\mathbf{y}}_a,{\mathbf{v}}_a,\sbbfa)$,
${\mathbf{J}}({\mathbf{y}}_a,{\mathbf{v}}_a,\sbbfa)$,
$\mathbf{G}(\mathbf{y}_a,\mathbf{v}_a,\sbbfa)$ for the ten conserved quantities,
and then (ii) to replace the harmonic-coordinate velocities
${\mathbf{v}}_a=\md{\mathbf{y}}_a/\md t$ in terms of Hamiltonian time-derivatives,
namely $\mathbf{V}_a=\{\mathbf{Y}_a,H\}$. Finally, the
values of the coefficients $a_1,\ldots,a_6$ and $b_1,\ldots,b_6$ must
fulfill the equations
\begin{subequations}
\label{iomcomp}
\begin{align}
E\big(\mathbf{Y}_a(\xpsb),\mathbf{V}_a(\xpsb),
\mathbf{\Sigma}_a(\xpsb)\big) &= H(\xpsa),
\\[1ex]
\mathbf{P}\big(\mathbf{Y}_a(\xpsb),\mathbf{V}_a(\xpsb),
\mathbf{\Sigma}_a(\xpsb)\big) &= \sum_a\mathbf{p}_a,
\\[0ex]
\mathbf{J}\big(\mathbf{Y}_a(\xpsb),\mathbf{V}_a(\xpsb),
\mathbf{\Sigma}_a(\xpsb)\big)
&= \sum_a\big(\mathbf{x}_a\times\mathbf{p}_a+\mathbf{S}_a\big),
\\[0ex]
\mathbf{G}\big(\mathbf{Y}_a(\xpsb),\mathbf{V}_a(\xpsb),
\mathbf{\Sigma}_a(\xpsb)\big) &= \mathbf{G}(\xpsa).
\end{align}
\end{subequations}
By considering the first three of these equations (i.e.\ by
comparing the two expressions for the energy, the total linear momentum,
and the total angular momentum), we obtained a unique set of values
for all the unknown coefficients $a_1,\ldots,a_6;b_1,\ldots,b_6$.
We then verified that these values satisfy also the fourth of Eqs.\ \eqref{iomcomp}
(thereby giving us confidence in the correctness of our Hamiltonian, and providing many
non-trivial checks of the previous results \cite{FBB06,BBF06}).

Our unique solution for the spin-dependent transformation of orbital coordinates
${\bf{Y}}^{\mathrm{so}}_{a(2)}+{\bf{Y}}^{\mathrm{so}}_{ a(4)}$ reads:
\begin{align}
\label{Y1expl}
{\bf{Y}}^{\mathrm{so}}_{1(2)}
({\mathbf{x}}_a,{\mathbf{p}}_a,{\mathbf{S}}_a)
+ {\bf{Y}}^{\mathrm{so}}_{1(4)}
({\mathbf{x}}_a,{\mathbf{p}}_a,{\mathbf{S}}_a)
&= \frac{\sipiv}{2c^2m_1^2}
-\frac{\sipiv}{c^4m_1^2}\Bigg(\frac{\pipi}{8m_1^2}
+\frac{Gm_2}{r_{12}}\Bigg)
\nonumber\\&\quad
+ \frac{G}{2c^4m_2r_{12}}
\Big(3\siipiiv + 2{\npii}\nsii + \siinpii{\mathbf{n}}_{12}
\Big).
\end{align}
Note that the first three terms on the right
side of Eq.\ \eqref{Y1expl} (i.e.\ the terms proportional to ${\mathbf{S}}_1$)
have the same structure as the exact special relativistic value 
\cite{Souriau,BelMartin,HansonRegge} for the
shift ${\bf{Y}}^{\mathrm{so}}_1$ between the canonical\footnote{The classical
canonical variables, here denoted
${\mathbf{x}}_a,{\mathbf{p}}_a,{\mathbf{S}}_a$, correspond,
at the quantum level, to the so-called Pryce-Newton-Wigner variables.}
orbital coordinate $\mathbf{x}_1$ and the usual Lorentz-covariant (harmonic)
orbital coordinate $\mathbf{y}_1$, namely
\be
\mathbf{y}_1 = \mathbf{x}_1 + \frac{\sipiv}{m_1 (m_1 c^2 + E_1)},
\ee
where $E_1= \sqrt{(m_1 c^2)^2 + ({\mathbf{p}}_1 c)^2}$ is the
relativistic energy (including the rest-mass contribution). The 
${\mathbf{p}}_1$-dependent terms in the first three terms
of Eq.\ \eqref{Y1expl} correspond to the
NLO expansion of the special-relativistic result, while the additional
$G$-dependent contribution can be roughly understood as a gravitational
addition to the special-relativistic energy $E_1$ (though it does not have
the correct coefficient to be really interpreted so simply).

By contrast, the terms proportional to ${\mathbf{S}}_2$ in Eq.\ \eqref{Y1expl}
do not have correspondants in the special-relativistic (i.e. $G \to 0$) limit.
As was to be expected they vanish in the limit where the second body (of mass $m_2$)
is heavy and fixed (${\mathbf{p}}_2/m_2 \to 0$). (Indeed, if we consider a non-spinning
test particle, $m_1$, $S_1=0$, moving in the background of a fixed, heavy spinning mass,
$m_2, S_2$, the harmonic-coordinate geodesic action of $m_1$ will already yield
a canonical Hamiltonian action.) We leave to future work a direct
derivation of these terms from the perturbative construction of
canonical coordinates (of the type $q = y + O(s)$, $p = p^{\rm bare} + O(s)$)
alluded to above.

Finally, as a further check on the algebra, we have also used the
``direct'' method (1) mentioned above (the first method we could have used 
to determine the values of the coefficients  $a_1,\ldots,a_6 ; b_1,\ldots,b_6$).
More explicitly, we started from the  harmonic-coordinate translational
equations of motion with NLO spin-orbit effects given
in Eqs.\  (5.3) of Ref.\ \cite{FBB06}. We then replaced in these
equations the non-conserved spin vector $\sbbfnca$ by its expression (as
given in Eq.\ (7.4) of \cite{BBF06}) in terms of their conserved spin vector
$\sbbfa$. This yields 2PN-accurate translational equations
of motion of the form
\begin{align}
\label{ahar}
\frac{\md\mathbf{v}_a}{\md t}
&= \mathbf{A}_{\mathrm{o}\,a}^{\mathrm{N}}(\mathbf{y}_b,\mathbf{v}_b)
+ \frac{1}{c^2} \Big(
\mathbf{A}_{\mathrm{o}\,a}^{\mathrm{1PN}}(\mathbf{y}_b,\mathbf{v}_b)
+ \mathbf{A}_{\mathrm{so}\,a}^{\mathrm{LO}}
(\mathbf{y}_b,\mathbf{v}_b,\sbbfb)
\Big)
\nonumber\\[1ex]&\quad + \frac{1}{c^4} \Big(
\mathbf{A}_{\mathrm{o}\,a}^{\mathrm{2PN}}(\mathbf{y}_b,\mathbf{v}_b)
+ \mathbf{A}_{\mathrm{so}\,a}^{\mathrm{NLO}}
(\mathbf{y}_b,\mathbf{v}_b,\sbbfb)
\Big) + {\cal O}(c^{-6}).
\end{align}
We then compared the right-hand-side of Eq.\ \eqref{ahar},
let us denote it by
$\mathbf{A}_a=\mathbf{A}_a(\mathbf{y}_b,\mathbf{v}_b,\sbbfb)$,
to its direct Hamiltonian recomputation by means of our Hamiltonian flow, i.e.\
\be
\label{aharequiv}
\mathbf{A}_a = \{\mathbf{V}_a,H\}
= \big\{\{\mathbf{Y}_a,H\},H\big\},
\ee
together with the needed transformations \eqref{transformations} (determined above)
between harmonic and canonical variables. Again, this verification
worked perfectly and (together with the similar direct verification
of the NLO spin precession equation mentioned above)
gives us confidence that {\it both} sets of results (harmonic and Hamiltonian) are correct.

\acknowledgments

This work is supported in part
by the KBN Grant no 1 P03B 029 27 (to P.J.)
and by the Deutsche Forschungsgemeinschaft (DFG) through
SFB/TR7 ``Gravitational Wave Astronomy.''
T.D.\ thanks ICRANet (Pescara, Italy) for partial support.

\end{document}